\documentclass{PoS}

\title{Upsilon polarization measurement at CDF}

\ShortTitle{Upsilon polarization measurement at CDF}

\author{\speaker{Thomas Kuhr} for the CDF collaboration\\
        Institut f\"ur Experimentelle Kernphysik, KIT, Karlsruhe, Germany\\
        E-mail: \email{Thomas.Kuhr@kit.edu}}

\abstract{Measurements of production cross sections and polarizations
are essential inputs for a theoretical understanding of heavy vector meson production.
In this article the CDF measurement of the $Y(1S)$ polarization in the kinematic
range $|y|<0.6$ and $2 < p_T$ [GeV$/c$] $< 40$ using a data sample of
$2.9$ fb$^{-1}$ is described.
Compared to the CDF Run I measurement, with which it agrees, it extends the covered $p_T$ range,
allowing for a better comparison to predictions in the perturbative regime.
The observed trend towards longitudinal polarization at high transverse momentum
disagrees with predictions based on non-relativistic QCD.}

\FullConference{XVIII International Workshop on Deep-Inelastic Scattering and Related Subjects\\
		 April 19 -23, 2010\\
		 Convitto della Calza, Firenze, Italy}

\begin{document}

\section{Introduction}
The understanding of heavy vector meson production at hadron colliders and $ep$ 
colliders has evolved in the past several years.
While the Color Singlet Model \cite{CSM}, describing the transition for the heavy 
quark anti-quark pair produced in the hard scattering process to the bound $Q\bar{Q}$ 
state by the emission of hard gluons, failed to reproduce the $J/\psi$ 
cross section measured at the Tevatron \cite{Abe:1997jz}, the addition of soft gluon emission in the 
Color Octet Model \cite{Kramer:2001hh} could solve this issue.
The theoretical framework usually applied here is non-relativistic QCD (NRQCD) \cite{Bodwin:1994jh}.
It factorizes the quarkonium creation into a hard process to produce the $Q\bar{Q}$ pair, 
and a long-distance matrix element for the meson formation.
As the non-perturbative matrix elements are believed to be universal, 
they can be measured in experiments and then applied in theoretical predictions.

One of the predictions of NRQCD is that the polarization of heavy vector mesons 
should be transverse at large transverse momenta.
Recent CDF measurements of the $J/\psi$ polarization \cite{Abulencia:2007us} disagree with this expectation.
A possible explanation for this discrepancy is that the charm quark is too light for perturbative calculations.
Fortunately, the Tevatron provides the possibility to test the predicted trend towards 
transverse polarization for the much heavier $Y$ states \cite{note}.

\section{Spin Alignment Measurement}
The polarization parameter $\alpha$ is defined as 
$\alpha = \frac{\sigma_T-2\sigma_L}{\sigma_T+2\sigma_L}$ where $\sigma_{T/L}$ are the 
cross sections for the production of transverse and longitudinal polarized vector mesons, respectively.
Thus the minimal and maximal values of $\alpha=-1$ and $\alpha=1$ correspond to full 
longitudinal and transverse  polarization, respectively. 
A value of $\alpha=0$ indicates unpolarized production.

The measurement of the spin alignment of the vector meson requires the definition of a reference axis.
In this measurement the reference axis is defined by the $Y$ momentum in the lab frame.
This definition, called s-channel helicity frame, is used by all polarization measurements at hadron colliders so far.

The polarization parameter $\alpha$ can be determined experimentally by measuring 
the distribution of the decay angle $\theta^*$ of the positively charged muon from the 
$Y \rightarrow \mu^+\mu^-$ decay in the $Y$ rest frame with respect to the reference axis.
The angular distribution is given by
$\frac{d\Gamma}{d(\cos\theta^*)} \propto 1 + \alpha \cos^2 \theta^*$.
Because $\cos\theta^*$ enters only quadratically, the distribution is symmetric about $0$.
This consequence of parity conservation allows to perform the measurement as function 
of $|\cos\theta^*|$ and thus to increase the statistics per bin in the angular distribution.

\section{Data Sample}
The analyzed data sample corresponds to an integrated luminosity of $2.9$ fb$^{-1}$ of
$p\bar{p}$ collisions at a center of mass energy of $\sqrt{s}=1.96$ TeV
recorded by the CDF II detector at the Tevatron collider.
The $Y$ events are triggered by a pair of oppositely charged muons with a transverse
momentum of $p_T(\mu) > 3$ GeV$/c$.
In the offline selection one muon is required in addition to have at least 
a transverse momentum of 4 GeV$/c$.
Only muons in the central region of the detector, with a pseudorapidity $|\eta(\mu)|<0.6$, are used.
Further selection criteria are applied to select well-reconstructed $Y$ mesons.
The kinematic range of $Y$ mesons studied in this analysis covers the central
rapitidy region, $|y| < 0.6$, and the transverse momentum range 
$2 < p_T$ [GeV$/c$] $< 40$.

\begin{figure}
\centering
\includegraphics[width=0.5\textwidth,clip=true]{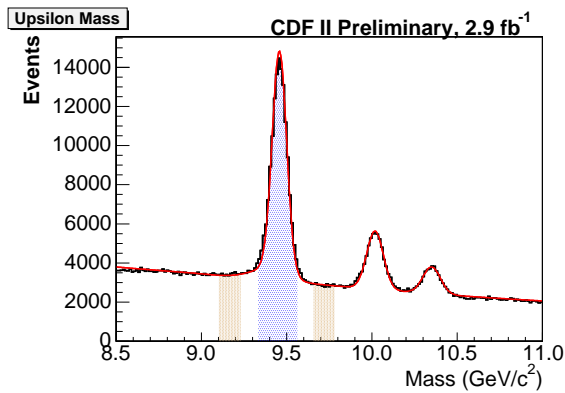}
\caption{Invariant $\mu^+\mu^-$ mass distribution with shaded $Y(1S)$ signal and side band regions.}
\label{fig:mass}
\end{figure}

Figure \ref{fig:mass} shows the invariant mass spectrum of the selected dimuon pairs
together with a fit where the signal shapes are taken from simulation and the
background is described by an exponential.
A range of $\pm2.5 \sigma$ around the $Y(1S)$ mass peak center is defined as signal region.
The good mass resolution allows to define lower and upper mass sidebands close to the signal
for the estimation of background events in the signal region based on the exponential background shape.
The selected data sample contains about 80.000 $Y(1S)$ mesons.

\section{Analysis Method and Systematic Uncertainties}
The same method to determine the polarization as used in Run I \cite{Acosta:2001gv} and 
for the Run II $\psi$ polarization measurement \cite{Abulencia:2007us} is applied here.
The data sample is split into bins of $p_T$ and then the $\cos\theta^*$ distribution
in each $p_T$ bin is examined.
A $\chi^2$ fit is performed to the $|\cos\theta^*|$ distributions with 10 equidistant bins.
The background contributions for all bins are free parameters in the fit and constrained
by simultaneously fitting the mass sideband regions.
The signal distribution is described by the weighted sum of two templates for transverse (T) and longitudinal (L)
polarization.
The relative weights of both contributions determine the polarization parameter $\alpha$.

\begin{figure}
\centering
\includegraphics[width=0.26\textwidth]{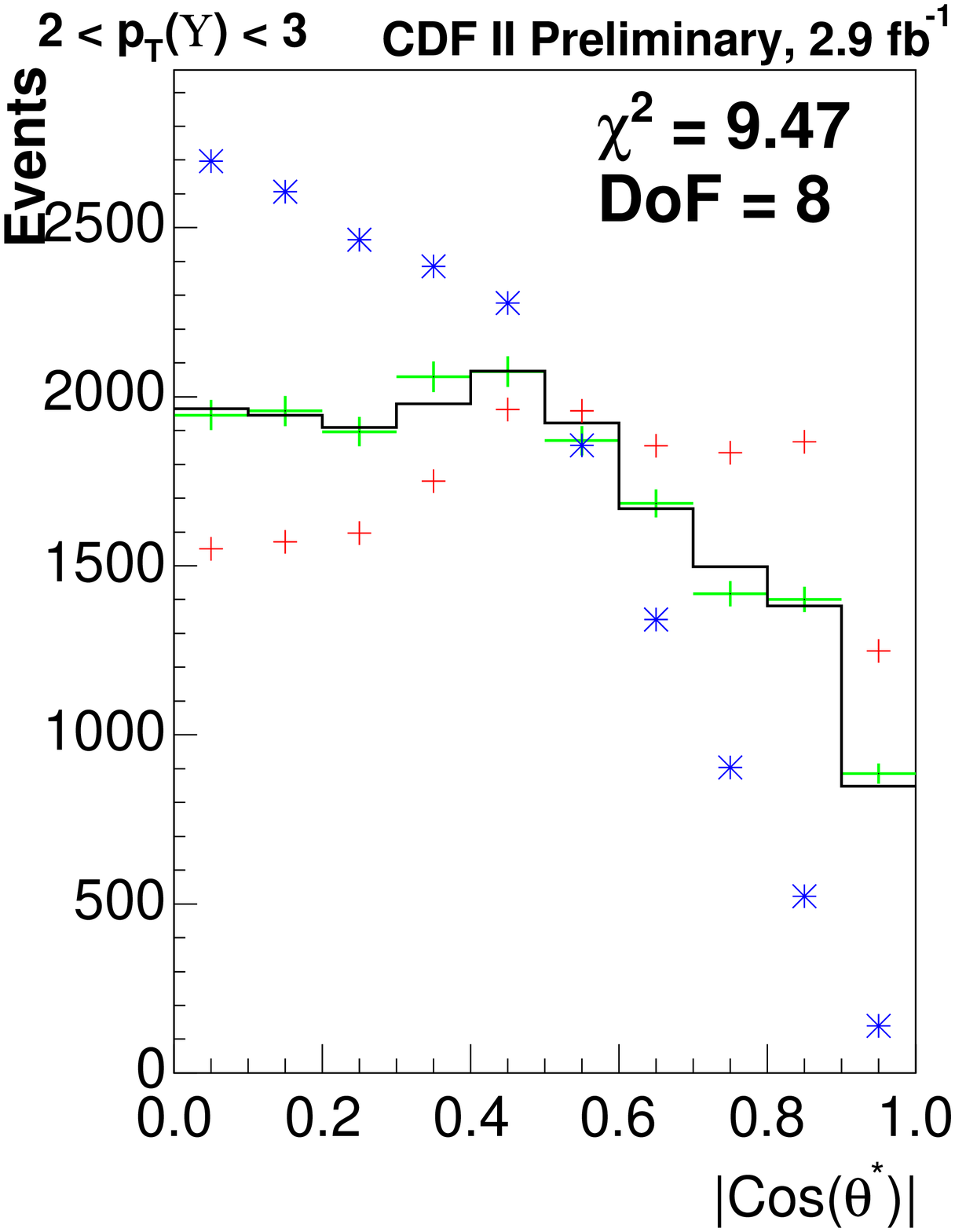}
\hspace*{0.02\textwidth}
\includegraphics[width=0.23\textwidth,clip,trim=0 206 430 0]{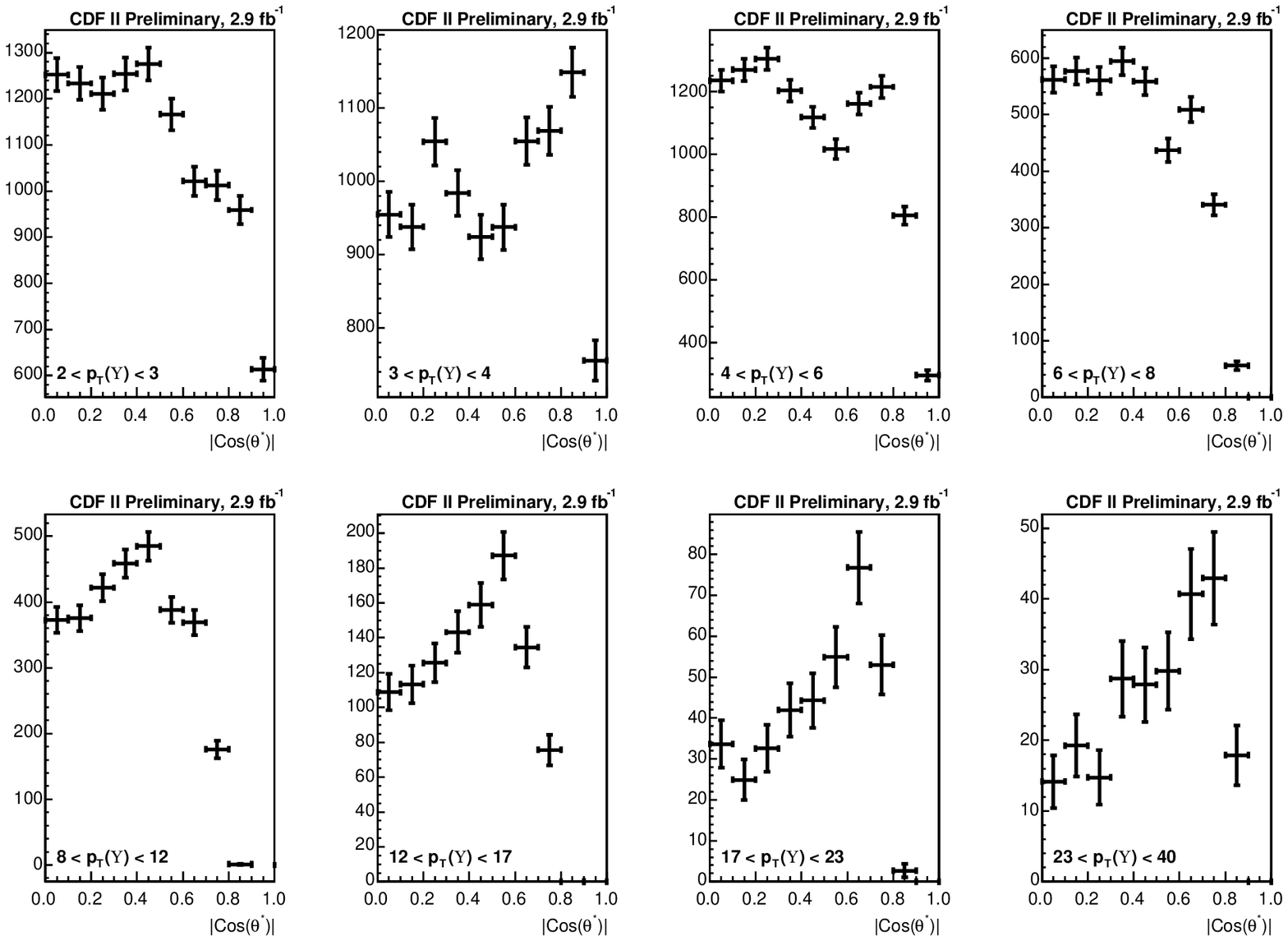}
\includegraphics[width=0.23\textwidth,clip,trim=0 0 430 206]{Y1S_bg.eps}
\includegraphics[width=0.23\textwidth,clip,trim=430 0 0 206]{Y1S_bg.eps}
\caption{Fit (black) to $|\cos\theta^*|$ distribution (green) for $2 < p_T$ [GeV$/c$] $<3$ with 
T (red) and L (blue) templates (left) and
fitted angular distributions of background events for selected $p_T$ bins (right three plots).}
\label{fig:fit}
\end{figure}

Because the $|\cos\theta^*|$ signal templates are affected by acceptance and efficiency effects,
they are determined from simulation.
Fully-transverse and fully-longitudinal polarized $Y(1S) \rightarrow \mu^+\mu^-$ samples were 
generated with the EvtGen package and simulated and reconstructed with the standard CDF software.
Trigger efficiency effects were taken into account by applying efficiencies measured on data
to the simulation.
Since the acceptance causes the $|\cos\theta^*|$ template distributions to be dependent on
the transverse momentum of the $Y$,
the simulated $p_T$ distribution is reweighted in an iterative procedure to match the one observed in data.
Figure \ref{fig:fit} shows the fit for one $p_T$ bin and the obtained background distributions
for low, medium, and high $p_T$.
The background shows a strong dependence on the transverse momentum.

Several systematic effects that could alter the $|\cos\theta^*|$ distributions were investigated.
The uncertainty on the trigger efficiency measured on data leads to a systematic uncertainty
of 0.007 on the polarization parameter.
Other systematic uncertainties were found to be negligible.
This includes variations of the mass shapes and the signal window definition, a change of the
$p_T(\mu)$ cut, and using an alternative $p_T$ reweighting.
Although the chosen bin width is much larger than the resolution in $\cos\theta^*$,
a check for a systematic effect was performed by using twice the number of bins.
No effect beyond expected statistical fluctuations was found.
It was also verified that there is no sizable migration between $p_T$ bins.
Overall the systematic uncertainties are much smaller than the statistical ones.

\section{Result}
\begin{figure}
\centering
\includegraphics[width=0.45\textwidth]{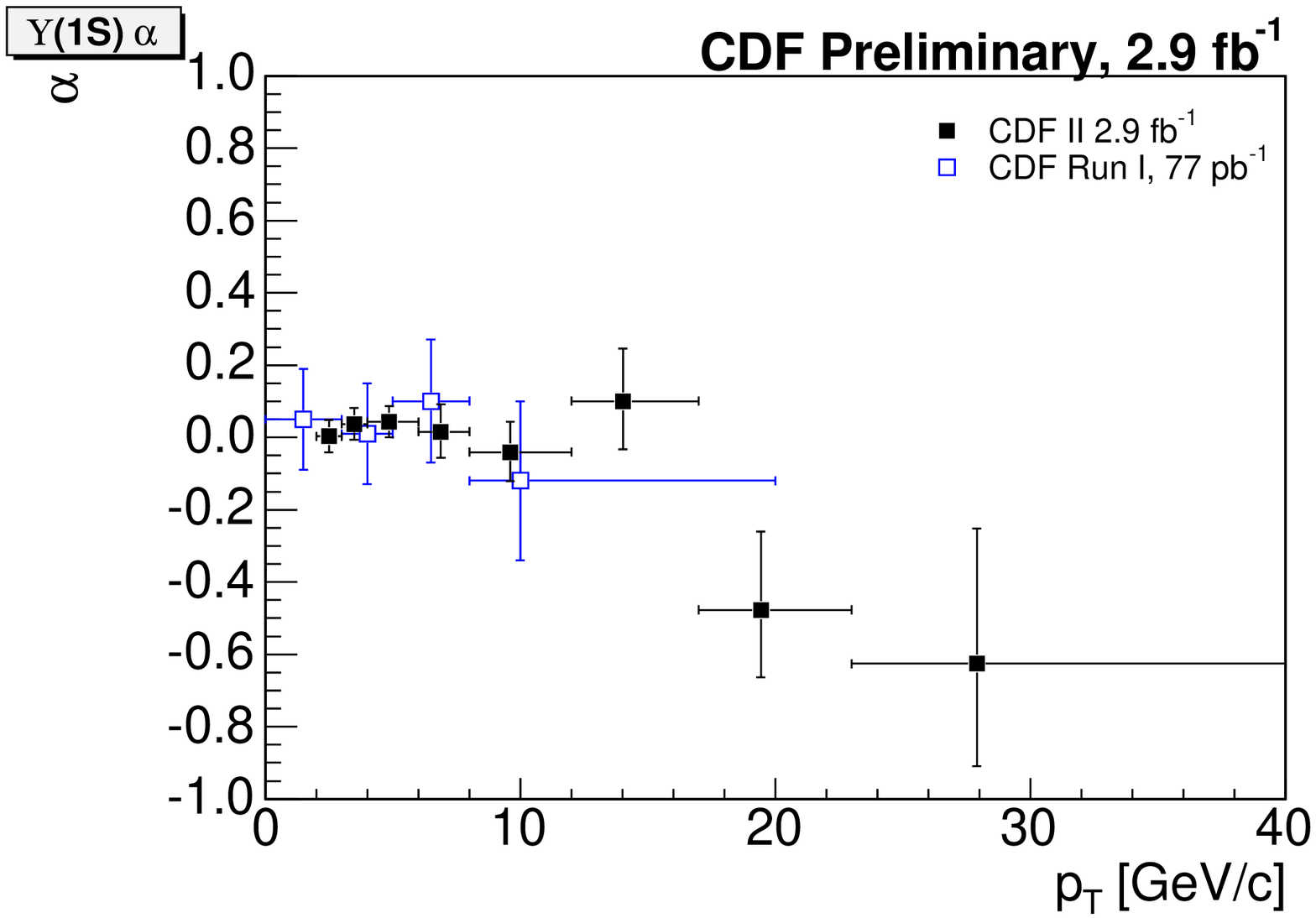}
\hspace*{0.05\textwidth}
\includegraphics[width=0.45\textwidth]{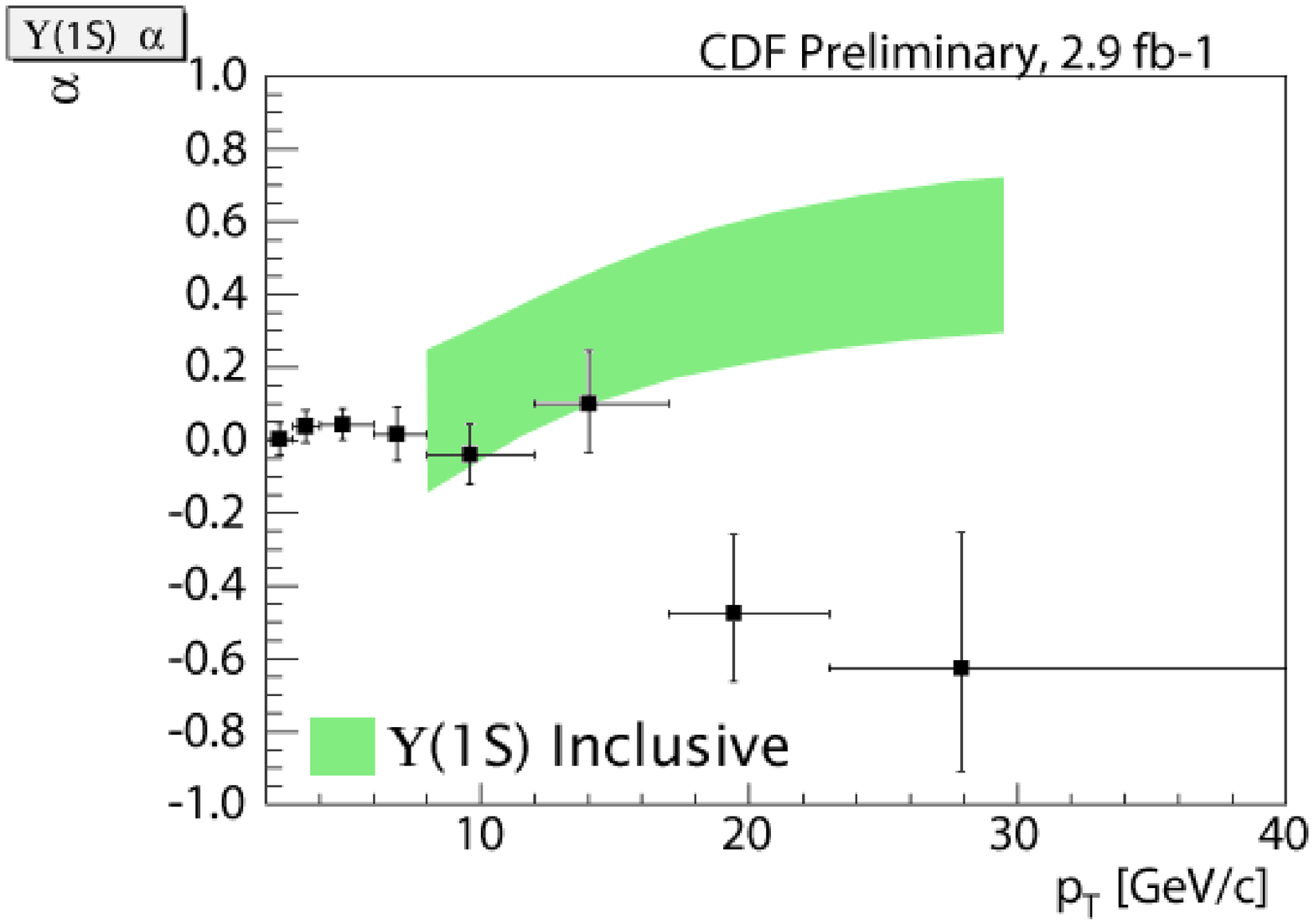}
\caption{$Y(1S)$ polarization measurement compared with the CDF Run I result (left) 
and a NRQCD prediction \cite{Braaten:2000gw} (right).}
\label{fig:result}
\end{figure}

Figure \ref{fig:result} shows the measured $Y(1S)$ polarization as a function of $p_T$.
It is compatible with zero polarization, with a trend towards longitudinal polarization
at high $p_T$.
As can be seen in the left plot, the result is consistent with the CDF Run I measurement 
\cite{Acosta:2001gv} and considerably extends its momentum range to higher $p_T$.
The right plot in Fig.~\ref{fig:result} illustrates that the measurement contradicts 
the NRQCD prediction of transverse polarization at high $p_T$.

\begin{figure}
\centering
\includegraphics[width=0.45\textwidth]{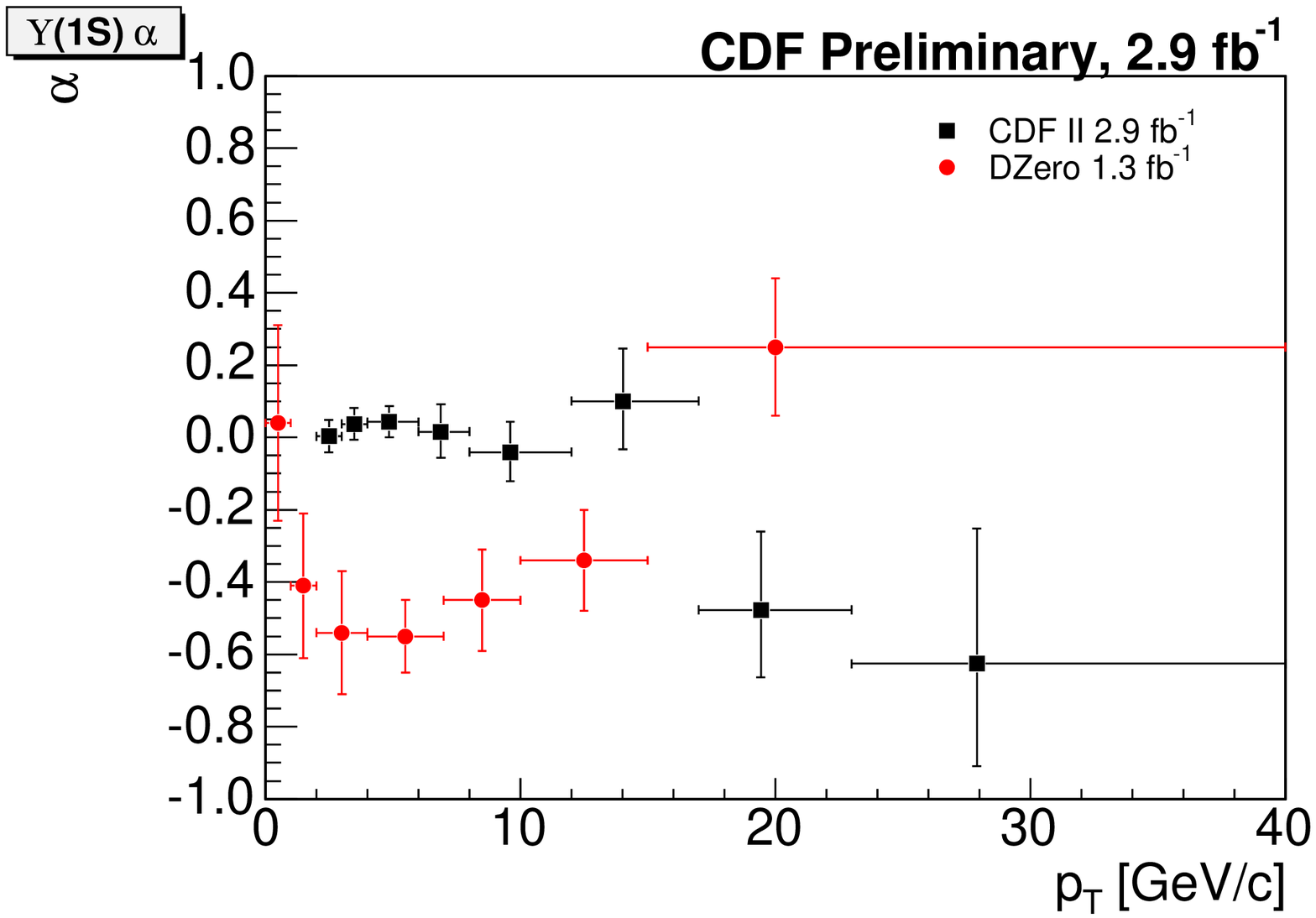}
\hspace*{0.05\textwidth}
\includegraphics[width=0.45\textwidth,clip,trim=0 0 300 190]{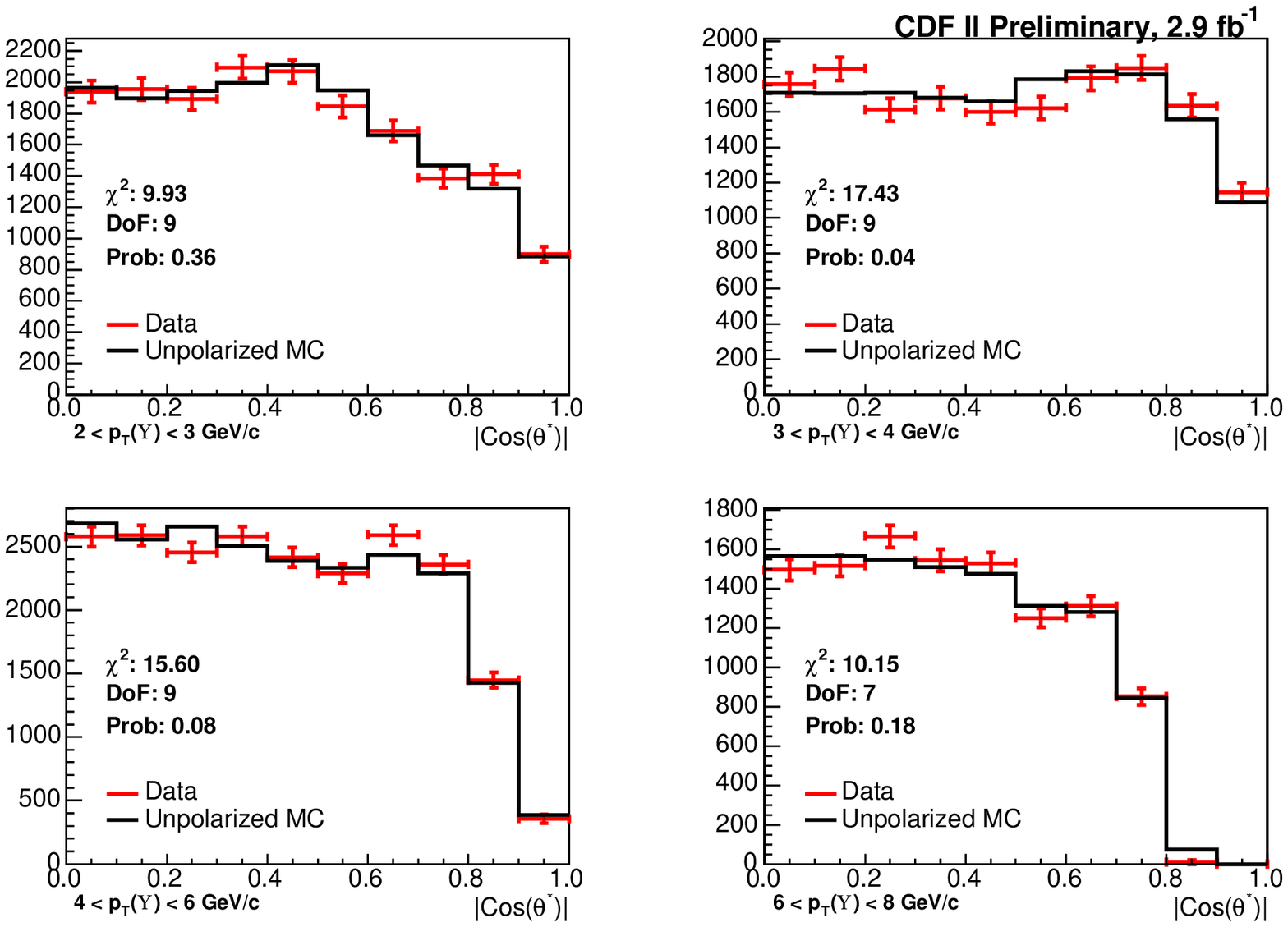}
\caption{$Y(1S)$ polarization measurement compared with the D0 result \cite{Abazov:2008za} (left) 
and comparison of data with simulated unpolarized events at low $p_T$ (right).}
\label{fig:comparison}
\end{figure}

As shown in Fig.~\ref{fig:comparison} the result disagrees with the D0 measurement \cite{Abazov:2008za}.
Since the D0 measurement indicates a longitudinal polarization at low $p_T$,
the result in this region was cross-checked by comparing the measured $|\cos\theta^*|$
distribution with the simulation of unpolarized events.
Good agreement is observed as shown exemplarily for one $p_T$ bin in the right plot in Fig.~\ref{fig:comparison},
confirming the consistency with zero polarization.

A possible reason for the discrepancy between the CDF and D0 results is that the 
D0 measurement was performed in a larger kinematic range of $|y|<1.7$.
Another difference between both measurements is caused by the different track reconstruction precisions.
Due to a worse mass resolution the signal shapes of $Y(1S)$, $Y(2S)$, and $Y(3S)$
overlap in the D0 measurement, not allowing to define mass sidebands as close to the signal
as in the CDF measurement.

\section{Summary and Outlook}
The $Y(1S)$ polarization is measured by CDF up to $p_T=40$ GeV$/c$.
It is consistent with zero and tends towards longitudinal polarization at high $p_T$.
The result is compatible with the CDF run I result and inconsistent with the D0 result.
It also disagrees with NRQCD predictions.

Updated measurements with more data, having data with about three time the integrated 
luminosity already recorded, can be expected, including polarization measurements of the $Y(2S)$ and $Y(3S)$.
This, together with a separation of direct and feed-down contributions, may help
to obtain a better theoretical understanding of heavy meson production.

\end{document}